\documentclass[12pt]{article}
\usepackage{type1cm}
\usepackage[dvipdfmx]{graphicx}
\usepackage{cite}

\usepackage{amsmath,amssymb,bm,amsthm,cases}

\def\beq{\begin{equation}}
\def\eeq{\end{equation}}
\def\nbeq{\begin{equation*}}
\def\neeq{\end{equation*}}

\def\<{\langle}
\def\>{\rangle}

\begin{document}
\title{\Large Weak eigenstate thermalization with large deviation bound}
\author{Takashi Mori \\
{\it
Department of Physics, Graduate School of Science,} \\
{\it The University of Tokyo, Bunkyo-ku, Tokyo 113-0033, Japan
}}
\date{}
\maketitle

\begin{abstract}
We investigate the eigenstate thermalization hypothesis (ETH) for a translationally invariant quantum spin system on the $d$-dimensional cubic lattice under the periodic boundary conditions.
It is known that the ETH holds in this model for typical energy eigenstates in the sense that the standard deviation of the expectation values of a local observable in the energy eigenstates within the microcanonical energy shell vanishes in the thermodynamic limit, which is called the weak ETH.
Here, it is remarked that the diagonal elements of a local observable in the energy representation shows the large deviation behavior.
This result implies that the fraction of atypical eigenstates which do not represent thermal equilibrium is exponentially small.
\end{abstract}

\section{Introduction}

The eigenstate thermalization hypothesis (ETH) is regarded as one of the promising perspectives to explain thermalization in isolated quantum systems~\cite{Srednicki1994,Deutsch1991,Rigol2008,Biroli2010,Tasaki2016_typicality}.
If \textit{every} energy eigenstate $|n\>$ of the Hamiltonian $H$ in the microcanonical energy shell $\mathcal{H}_{E,\Delta E}:=\mathrm{span}\{|n\>: H|n\>=E_n|n\>, E_n\in[E-\Delta E,E]\}$ is locally indistinguishable from the microcanonical ensemble $\rho^{\mathrm{mc}}$ in the sense that $\<n|O|n\>\approx\mathrm{Tr}\, O\rho^{\mathrm{mc}}$ for any local observable $O$ and large system size $N$ ($N$ is the number of lattice sites in a spin system), the Hamiltonian $H$ is said to satisfy the (strong) ETH.
There is no rigorous proof of the ETH although it is believed that the ETH holds at least for a large class of nonintegrable systems.
On the other hand, a weaker version of the ETH (weak ETH), which states that \textit{almost all} energy eigenstates $|n\>\in\mathcal{H}_{E,\Delta E}$ are locally indistinguishable from the microcanonical ensemble, is known to hold for general translationally invariant short-range interacting systems~\cite{Biroli2010}.
It is shown that the standard deviation of $\<n|O|n\>$ under the uniform distribution on $n$ with $|n\>\in\mathcal{H}_{E,\Delta E}$ is polynomially small in $N$ (in many cases $N^{-1/2}$), which implies, by applying the Chebyshev inequality, that the fraction of non-thermal energy eigenstates is at least polynomially small in $N$.

Here, we consider a translationally invariant quantum spin system on the $d$-dimensional cubic lattice under the periodic boundary conditions, and it is remarked that one can show that the fraction of non-thermal energy eigenstates is \textit{exponentially} small in $N$ for many cases.
This large deviation type bound on the probability of the fluctuation of $\<n|O|n\>$ is much stronger than the usual statement of the weak ETH relying on the Chebyshev inequality.

\section{Setting and weak ETH}

Let us consider a quantum spin system on the $d$-dimensional cubic lattice $\Lambda$ of side $L$ with the periodic boundary conditions.
The number of sites in $\Lambda$ is denoted by $N=L^d$.
We assume the translational invariance of the Hamiltonian $H$.
The normalized $n$th energy eigenstate is denoted by $|n\>$, i.e., $H|n\>=E_n|n\>$.
The whole Hilbert space is denoted by $\mathcal{H}$.
As in Introduction, the microcanonical energy shell is denoted by $\mathcal{H}_{E,\Delta E}$ and the microcanonical density matrix is defined as $\rho^{\mathrm{mc}}_N:=\mathsf{P}_{\mathcal{H}_{E,\Delta E}}/\mathrm{dim}\, \mathcal{H}_{E,\Delta E}$, where $\mathsf{P}_{\mathcal{H}_{E,\Delta E}}$ is the projection to $\mathcal{H}_{E,\Delta E}$.
The microcanonical average of an operator $A$ is denoted by $\<A\>^{\mathrm{mc}}_N:=\mathrm{Tr}\, \rho^{\mathrm{mc}}_NA$.

For a local observable $O$ with a finite support, let us consider its diagonal elements in the energy representation $\<n|O|n\>$ with $|n\>\in\mathcal{H}_{E,\Delta E}$.
The statement of the weak ETH is expressed as follows: for any $\delta>0$,
\beq
\mathrm{Prob}\{|\<n|O|n\>-\<O\>^{\mathrm{mc}}_N|>\delta\}\rightarrow 0 \text{ as } N\rightarrow\infty,
\eeq
where $\mathrm{Prob}\{\chi\}$ is the probability of $\chi$ in the uniform distribution of $n$ with $|n\>\in\mathcal{H}_{E,\Delta E}$.
In Ref.~\cite{Biroli2010}, it is argued that the weak ETH is a general property of translationally invariant quantum spin systems by using the Chebyshev inequality.

Here we consider the weak ETH with large deviation bound, which states that for any given $\delta>0$, there exists $\gamma>0$ such that
\beq
\mathrm{Prob}\{|\<n|O|n\>-\<O\>^{\mathrm{mc}}_N|>\delta\}\leq e^{-N\gamma}.
\label{eq:LD_ETH}
\eeq
This inequality implies that the fraction of non-thermal energy eigenstates is exponentially small in $N$.

We will see in Secs.~\ref{sec:main} and \ref{sec:proof} that (\ref{eq:LD_ETH}) is proved for many cases including short-range interacting Hamiltonian $H$ in the whole range of the energy corresponding to positive temperature for $d=1$ and in the range of the energy corresponding to high temperatures for $d\geq 2$.

\section{Large deviation upper bound for macro observables}
\label{sec:large}

For a local observable $O$, we consider the macrovariable $M$ defined as the translationally invariant sum of $O$:
\beq
M:=\frac{1}{N}\sum_{x\in\Lambda}O_x.
\eeq
We assume that $M$ is bounded, and hence $m_0:=\|M\|<+\infty$, where $\|\cdot\|$ denotes the operator norm.

Let us introduce the ``rate function'' $I(m)$ defined as the maximum lower semicontinuous function such that
\beq
\limsup_{N\rightarrow\infty}\frac{1}{N}\ln\left\<\mathsf{P}(M\in\Gamma)\right\>^{\mathrm{mc}}_N \leq -\inf_{m\in\bar{\Gamma}}I(m)
\label{eq:rate}
\eeq
for any interval $\Gamma\subset [-m_0,m_0]$, where $\mathsf{P}(M\in\Gamma)$ is the projection operator onto the Hilbert subspace, $\mathrm{span}\{|\phi\>\in\mathcal{H}: M|\phi\>=m|\phi\>, m\in\Gamma\}$, and $\bar{\Gamma}$ is the closure of $\Gamma$.
The right hand side of (\ref{eq:rate}) is called the large deviation upper bound\footnote{$I(m)$ is the rate function in the large deviation principle if the large deviation lower bound 
$$\liminf_{N\rightarrow\infty}\frac{1}{N}\ln\left\<\mathsf{P}(M\in\Gamma)\right\>^{\mathrm{mc}}_N\geq -\inf_{m\in\Gamma^o}I(m)$$
is also satisfied, where $\Gamma^o$ is the interior of $\Gamma$~\cite{Dembo_text}.}.
The rate function should be nonnegative $I(m)\geq 0$ and there exists some $m^*$ with $I(m^*)=0$.
Without loss of generality, we can choose $m^*=0$ and thus $I(0)=0$.

It can be shown that
\beq
\limsup_{N\rightarrow\infty}\frac{1}{N}\ln\left\<e^{N\lambda M}\right\>^{\mathrm{mc}}_N\leq \phi(\lambda):=\sup_m[\lambda m-I(m)]
\label{eq:Varadhan}
\eeq
for $\lambda\in\mathbb{R}$, which is a part of Varadhan's theorem~\cite{Dembo_text}.
The function $\phi(\lambda)$ is the Legendre-Fenchel transform of $I(m)$, and thus $\phi(\lambda)$ is convex.
It is noted that $\phi(0)=-\inf_m I(m)=0$ and $\phi(\lambda)\geq 0$.
Moreover, $\phi(\lambda)$ is a non-decreasing (non-increasing) function of $\lambda$ for positive (negative) $\lambda$.

For self-containedness, the proof of (\ref{eq:Varadhan}) is given in Appendix~\ref{appendix}.

In translationally invariant quantum spin systems with short-range interactions, it has been shown that $I(m)>0$ for any $m\neq 0$ and $I(m)$ is convex, which implies $I(m)=I^{**}(m)$ in the whole range of the energy corresponding to positive temperatures for $d=1$ and in the range of the energy corresponding to high temperatures for $d\geq 2$~\cite{Netocny2004,Lenci2005,Ogata2010}~\footnote{Originally, this has been shown for the canonical ensemble, but we can also show the same large deviation upper bound in the microcaonical ensemble, see Sec. 8 of Ref.~\cite{Tasaki2016_typicality}. In Ref.~\cite{Mori2016_macrostate}, $I_{\mathrm{mc}}(m)\geq I_{\mathrm{can}}(m)$ is explicitly shown ($I_{\mathrm{mc}}$ and $I_{\mathrm{can}}$ are the rate functions in the microcanonical ensemble and the canonical ensemble, respectively).}.

\section{Main result}
\label{sec:main}

Let us consider a local observable $O$ and the corresponding macro observable $M=(1/N)\sum_{x\in\Lambda}O_x$.
Here, $I(m)$ is the rate function for $M$ in the sense of (\ref{eq:rate}), and without loss of generality, we can put $I(0)=0$.
The function $\phi(\lambda)$ is defined by (\ref{eq:Varadhan}).
In order to show the weak ETH with large deviation bound (\ref{eq:LD_ETH}), we prove the following large deviation upper bound for the diagonal elements of a local observable $O$:
\beq
\mathrm{Prob}\{\<n|O|n\>\in\Gamma\}\leq e^{-N\inf_{m\in\bar{\Gamma}}I^{**}(m)+o(N)},
\label{eq:main}
\eeq
where $I^{**}(m)$ is the Legendre-Fenchel transform of $\phi(\lambda)$:
\beq
I^{**}(m):=\sup_{\lambda}[\lambda m-\phi(\lambda)].
\eeq
The function $I^{**}(m)$ is obtained by performing the Legendre-Fenchel transform twice on $I(m)$, and hence $I^{**}(m)$ is nothing but the convex envelope of $I(m)$.

This result is roughly expressed as follows.
If we have the following large deviation upper bound for a macrovariable $M=(1/N)\sum_{x\in\Lambda}O_x$:
\beq
P_{\mathrm{mc}}(M\approx m):=\<\mathsf{P}(M\approx m)\>^{\mathrm{mc}}_N \lesssim e^{-NI(m)},
\eeq
then the diagonal elements of the local operator $\<n|O|n\>$ has the large deviation upper bound with the rate function $I^{**}(m)$:
\beq
\mathrm{Prob}\left\{\<n|O|n\>\approx m\right\}\lesssim e^{-NI^{**}(m)}.
\eeq

From (\ref{eq:main}), we have, for any given $\delta>0$,
\beq
\mathrm{Prob}\left\{|\<n|O|n\>|>\delta\right\}\leq e^{-N\min\{I^{**}(\delta),I^{**}(-\delta)\}+o(N)}
=: e^{-N\gamma+o(N)}
\eeq
with $\gamma\geq 0$.
If $\gamma>0$, this proves the weak ETH with large deviation bound (\ref{eq:LD_ETH}): the fraction of non-thermal energy eigenstates with $|\<n|O|n\>-\<O\>^{\mathrm{mc}}_N|>\delta$ is exponentially small in the system size $N$ (note that $\<O\>^{\mathrm{mc}}_N=0$ in our choice).
This large deviation bound strengthens the usual statement of the weak ETH.
It is noted that $\gamma>0$ is equivalent to the condition that $I(m)$ attains a unique minimum.

As was explained in Sec.~\ref{sec:large}, it has been shown that $I(m)=I^{**}(m)>0$ for any $m\neq 0$ and thus $\gamma>0$ for short-range interacting systems in the whole range of the energy corresponding to positive temperatures for $d=1$ and in the range of the energy corresponding to high temperatures for $d\geq 2$~\cite{Netocny2004,Lenci2005,Ogata2010}.
Therefore, in these cases, the weak ETH with large deviation bound is proved.

For other cases such as long-range interacting systems and short-range interacting systems with $d\geq 2$ in an intermediate energy region, the main result (\ref{eq:main}) is still valid, but in order to conclude (\ref{eq:LD_ETH}), we must assume $\gamma>0$ for a small $\delta$.

\section{Proof}
\label{sec:proof}

Now we prove (\ref{eq:main}).
The proof is done in a standard way using the Markov inequality.

First, for an interval $\Gamma$, its closure is written as $\bar{\Gamma}=[m_1,m_2]$ with some $m_1$ and $m_2$ with $-m_0\leq m_1<m_2\leq m_0$.
Thus we have
\begin{align}
\mathrm{Prob}\left\{\<n|O|n\>\in\Gamma\right\}
&\leq\mathrm{Prob}\left\{\<n|O|n\>\in\bar{\Gamma}\right\}
\nonumber \\
&\leq \min\left\{\mathrm{Prob}\{\<n|O|n\>\geq m_1\}, \mathrm{Prob}\{\<n|O|n\>\leq m_2\}\right\}.
\label{eq:prob}
\end{align}
First we consider $\mathrm{Prob}\{\<n|O|n\>\geq m_1\}$.
Since 
\beq
\<n|O|n\>=\frac{1}{N}\sum_{x\in\Lambda}\<n|O_x|n\>=\<n|M|n\>
\eeq
because of the translational invariance, we have for $\lambda>0$,
\begin{align}
\mathrm{Prob}\{\<n|O|n\>\geq m_1\}
&=\mathrm{Prob}\left\{e^{N\lambda\<n|M|n\>}\geq e^{N\lambda m_1}\right\}
\nonumber \\
&\leq\frac{1}{\mathrm{dim}\, \mathcal{H}_{E,\Delta E}}\sum_{n:|n\>\in\mathcal{H}_{E,\Delta E}}e^{-N\lambda m_1}e^{N\lambda\<n|M|n\>}.
\end{align}
By using $e^{N\lambda\<n|M|n\>}\leq \<n|e^{N\lambda M}|n\>$, we have
\begin{align}
\mathrm{Prob}\{\<n|O|n\>\geq m_1\}&\leq e^{-N\lambda m_1}\frac{1}{\mathrm{dim}\, \mathcal{H}_{E,\Delta E}}\sum_{n:|n\>\in\mathcal{H}_{E,\Delta E}}\<n|e^{N\lambda M}|n\>
\nonumber \\
&=e^{-N\lambda m_1}\left\<e^{N\lambda M}\right\>^{\mathrm{mc}}_N.
\end{align}
Equation (\ref{eq:Varadhan}) implies
\beq
\left\<e^{N\lambda M}\right\>^{\mathrm{mc}}_N\leq e^{N\phi(\lambda)+o(N)},
\eeq
and therefore
\beq
\mathrm{Prob}\{\<n|O|n\>\geq m_1\}\leq e^{-N[\lambda m_1-\phi(\lambda)]+o(N)}.
\eeq
Because $\lambda>0$ is arbitrary, we have
\beq
\mathrm{Prob}\{\<n|O|n\>\geq m_1\}\leq e^{-N\sup_{\lambda>0}[\lambda m_1-\phi(\lambda)]+o(N)}.
\label{eq:lower}
\eeq

When $m_1>0$, since $\lambda m_1-\phi(\lambda)$ is an increasing function of $\lambda$ for $\lambda\leq 0$, 
\beq
\sup_{\lambda>0}[\lambda m_1-\phi(\lambda)]=\sup_{\lambda}[\lambda m_1-\phi(\lambda)]=I^{**}(m_1).
\eeq
Since $I^{**}(m)$ is convex, $I^{**}(0)=0$, $I^{**}(m)\geq 0$, and thus $I^{**}(m)$ is non-decreasing for $m>0$, we can write $I^{**}(m_1)=\inf_{m\geq m_1}I^{**}(m)$.

When $m_1\leq 0$, $\lambda m_1-\phi(\lambda)$ is a non-increasing function for $\lambda>0$, and hence $\sup_{\lambda>0}[\lambda m_1-\phi(\lambda)]=-\phi(0)=0$.
Since $I^{**}(0)=0$ and $I^{**}(m)\geq 0$, we have $0=\inf_{m\geq m_1}I^{**}(m)$.

Thus for any $m_1$, we obtain
\beq
\sup_{\lambda>0}[\lambda m_1-\phi(\lambda)]=\inf_{m\geq m_1}I^{**}(m).
\eeq
By substituting it into (\ref{eq:lower}), we obtain
\beq
\mathrm{Prob}\{\<n|O|n\>\geq m_1\}\leq e^{-N\inf_{m\geq m_1}I^{**}(m)+o(N)}.
\eeq

In a similar manner (by choosing $\lambda<0$), we can also show
\beq
\mathrm{Prob}\{\<n|O|n\>\leq m_2\}\leq e^{-N\inf_{m\leq m_2}I^{**}(m)+o(N)}.
\eeq

From (\ref{eq:prob}), we obtain
\begin{align}
\mathrm{Prob}\{\<n|O|n\>\in\Gamma\}
&\leq\min\left\{e^{-N\inf_{m\geq m_1}I^{**}(m)+o(N)},e^{-N\inf_{m\leq m_2}I^{**}(m)+o(N)}\right\}
\nonumber \\
&=e^{-N\inf_{m\in[m_1,m_2]}I^{**}(m)+o(N)}
\nonumber \\
&=e^{-N\inf_{m\in\bar{\Gamma}}I^{**}(m)+o(N)}.
\end{align}
This is the desired upper bound.

\section{Discussion}

For translationally invariant $d$-dimensional quantum spin systems, the weak ETH with large deviation bound for a local observable $O$ has been proved under the assumption that the rate function $I(m)$ of the macroscopic intensive observable $M=(1/N)\sum_{x\in\Lambda}O_x$ attains a unique minimum.
This assumption has been proved to hold for short-range interacting systems in the whole range of the energy corresponding to positive temperatures ($d=1$) and in some range of the energy corresponding to high temperatures ($d\geq 2$).

From the weak ETH with large deviation bound, we can obtain a sufficient condition of thermalization in isolated quantum systems.
Let $|\psi(0)\>=\sum_{n:|n\>\in\mathcal{H}_{E,\Delta E}}c_n|n\>$ be a normalized initial state.
After the time evolution under the Hamiltonian $H$, the state at time $t$ is given by $|\psi(t)\>=\sum_{n:|n\>\in\mathcal{H}_{E,\Delta E}}c_ne^{-iE_nt/\hbar}|n\>$, and the infinite-time average of $|\psi(t)\>\<\psi(t)|$ is given by $\rho_D=\sum_{n:|n\>\in\mathcal{H}_{E,\Delta E}}|c_n|^2|n\>\<n|$, where we have assumed that there is no energy degeneracy.
Equation~(\ref{eq:LD_ETH}) implies that for a given small $\delta>0$ and sufficiently large $N$,
\beq
\left|\mathrm{Tr}\, \rho_DO-\<O\>^{\mathrm{mc}}_N\right|<\delta \text{ if } D_{\mathrm{eff}}>e^{-\eta N}\mathrm{dim}\, \mathcal{H}_{E,\Delta E}
\eeq
with some $\eta\in(0,\gamma)$.
Here, the effective dimension $D_{\mathrm{eff}}$ is given by the inverse participation ratio, $D_{\mathrm{eff}}=\left(\sum_{n:|n\>\in\mathcal{H}_{E,\Delta E}}|c_n|^4\right)^{-1}$~\cite{Tasaki2016_typicality}, or $D_{\mathrm{eff}}=e^{S(\rho_D)}$~\cite{Mori2016_macrostate}, where $S(\rho)$ is the von Neumann entropy of $\rho$.

It is also shown that the temporal fluctuation of $\<\psi(t)|O|\psi(t)\>$ is very small~\cite{Tasaki1998,Reimann2008,Short2011,Tasaki2016_typicality}, and hence, not only the long-time average, but also $\<\psi(t)|O|\psi(t)\>$ is very close to $\<O\>^{\mathrm{mc}}_N$ for typical times $t$.
This means that thermalization can be proved even if the effective dimension is exponentially smaller than the dimension of the microcanonical energy shell.

Since the above result is valid even for integrable systems, the absence of thermalization reported in integrable systems implies that $D_{\mathrm{eff}}\ll e^{-\gamma N}\mathrm{dim}\,\mathcal{H}_{E,\Delta E}$ in quantum quench.
On the other hand, numerical calculations for non-integrable systems seem to suggest that $D_{\mathrm{eff}}\approx\mathrm{dim}\,\mathcal{H}_{E,\Delta E}$~\cite{Rigol2016}.
We should understand the reason of this difference in the behavior of effective dimensions in integrable and non-integrable systems, which is an open problem.

\section*{Acknowledgements}
I would like to thank Takahiro Sagawa and Hal Tasaki for fruitful discussion.

\appendix
\section{Proof of (\ref{eq:Varadhan})}
\label{appendix}

In this Appendix, we shall prove (\ref{eq:Varadhan}).
We assume $\lambda\geq 0$, but the proof for $\lambda<0$ can be done similarly.
First we decompose the Hilbert space by using
\beq
\hat{1}=\sum_{k=1}^K\mathsf{P}(M\in[m_k,m_{k+1})),
\eeq
where the left hand side is the identity operator on $\mathcal{H}$, $m_k=-m_0+(k-1)\Delta m$, and $K$ is the minimum integer greater than $2m_0/\Delta m$ with a given $\Delta m>0$.
By using this decomposition, for $\lambda\geq 0$,
\begin{align}
\left\<e^{N\lambda M}\right\>^{\mathrm{mc}}_N
&=\sum_{k=1}^K\left\<\mathsf{P}(M\in[m_k,m_{k+1}))e^{N\lambda M}\right\>^{\mathrm{mc}}_N
\nonumber \\
&\leq\sum_{k=1}^Ke^{N\lambda m_{k+1}}\left\<\mathsf{P}(M\in[m_k,m_{k+1}))\right\>^{\mathrm{mc}}_N
\nonumber \\
&\leq K\exp\left\{N\max_{k=1}^K\left[\lambda m_{k+1}+\frac{1}{N}\ln\<\mathsf{P}(M\in[m_k,m_{k+1}))\>^{\mathrm{mc}}_N\right]\right\}.
\end{align}
Thus we have
\begin{align}
\limsup_{N\rightarrow\infty}\frac{1}{N}\ln\left\<e^{N\lambda M}\right\>^{\mathrm{mc}}_N
&\leq \max_{k=1}^K\left[\lambda m_{k+1}+\limsup_{N\rightarrow\infty}\frac{1}{N}\ln\<\mathsf{P}(M\in[m_k,m_{k+1}))\>^{\mathrm{mc}}_N\right]
\nonumber \\
&\leq \max_{k=1}^K\left[\lambda m_{k+1}-\inf_{m\in[m_k,m_{k+1}]}I(m)\right].
\end{align}
Since $I(m)$ is a lower semicontinuous function, $\inf_{m\in[m_k,m_{k+1}]}I(m)=\min_{m\in[m_k,m_{k+1}]}I(m)=I(\tilde{m}_k)$ with some $\tilde{m}_k\in[m_k,m_{k+1}]$.
We also have $\lambda m_{k+1}\leq \lambda\tilde{m}_k+\lambda\Delta m$, and hence
\begin{align}
\limsup_{N\rightarrow\infty}\frac{1}{N}\ln\left\<e^{N\lambda M}\right\>^{\mathrm{mc}}_N
&\leq\max_{k=1}^K\left[\lambda\tilde{m}_k-I(\tilde{m}_k)\right]+\lambda\Delta m
\nonumber \\
&\leq\sup_{m}\left[\lambda m-I(m)\right]+\lambda\Delta m
\nonumber \\
&=\phi(\lambda)+\lambda\Delta m.
\end{align}
Since $\Delta m>0$ can be arbitrarily small, we obtain
\beq
\limsup_{N\rightarrow\infty}\frac{1}{N}\ln\left\<e^{N\lambda M}\right\>^{\mathrm{mc}}_N\leq\phi(\lambda).
\eeq
We can also prove this inequality for $\lambda<0$ similarly.

\end{document}